\documentclass[%
10pt,A4paper,twocolumn,aps,prl, reprint, superscriptaddress,longbibliography, nofootinbib,
 amsmath,amssymb,
]{revtex4-2}
\usepackage{lipsum,etoolbox}
\usepackage{mathtools}
\usepackage{graphicx}
\usepackage{dcolumn}
\usepackage{bm}
\usepackage{xcolor}
\usepackage{siunitx}
\usepackage{booktabs} 
\usepackage{array}
\usepackage{calrsfs}
\usepackage{multirow}
\usepackage{makecell}
\usepackage{enumitem}
\usepackage{placeins}
\usepackage{comment}
\usepackage{upgreek}
\usepackage[shortcuts]{extdash}
\usepackage[colorlinks=true, linkcolor=blue, citecolor=black, urlcolor=blue, pdfauthor={Yongxin Song}, pdftitle={microwave_crosstalk}]{hyperref}

\setlist[itemize]{leftmargin=*}

\begin{document}
\renewcommand{\arraystretch}{1.1}

\preprint{APS/123-QED}

\title{Microwave Crosstalk in Planar Superconducting Quantum Devices}

\author{Yongxin~Song}
\email{yongxin.song@phys.ethz.ch}
\author{Dominic~Hagmann}
\author{Kieran~Dalton}
\author{Felix~Henrich}
\author{Felix~Wagner}
\author{Mohsen~Bahrami~Panah}
\author{Marek~Pechal}
\author{Andreas~Wallraff}

\affiliation{Department of Physics, ETH Zurich, CH-8093 Zurich, Switzerland}
\affiliation{Quantum Center, ETH Zurich, CH-8093 Zurich, Switzerland}
\affiliation{ETH Zurich - PSI Quantum Computing Hub, Paul Scherrer Institute, CH-5232 Villigen, Switzerland}

\begin{abstract}
Microwave crosstalk poses a major challenge to scaling superconducting quantum devices as it introduces excess control errors. 
Although its magnitude and impact have been explored in various experimental settings, quantitative physical models capable of explaining measured crosstalk for a given device geometry remain scarce.
Here, we address this gap by investigating microwave crosstalk in planar superconducting devices with crossovers.
We identify two structures that can lead to strong crosstalk: a drive line routed in close proximity to another qubit, and a drive line crossing a qubit-qubit coupler using an air bridge.
We design and characterize devices involving these structures and develop physical models that quantitatively explain the experimentally observed crosstalk. 
Based on these models, we discuss the design considerations for reducing microwave crosstalk.
Our results provide practical guidance for low-crosstalk device layouts and establish a basis for the systematic investigation of weaker crosstalk mechanisms.
\end{abstract}

\date{\today}

\maketitle
\setcounter{secnumdepth}{3}

\section{Introduction}

Realizing large-scale quantum computation requires not only increasing the number of qubits, but also maintaining precise and selective control over increasingly complex quantum hardware. 
In superconducting quantum circuits, a major obstacle appears as signal crosstalk, which causes control errors~\cite{Gambetta2012, Patterson2019, Rudinger2021b, Ketterer2023, Santos2023} and becomes increasingly severe in large-scale devices due to frequency crowding~\cite{Morvan2022a, Osman2023} and the growing number of parasitic coupling pathways~\cite{Kosen2024}. 
Crosstalk-induced errors can be mitigated at the control level, for example by applying the inverse of the crosstalk-induced operation~\cite{Winick2021, Dai2021a, Nuerbolati2022, Wang2022q, Barrett2023} or by using crosstalk-robust control pulses~\cite{Wesdorp2026}. 
However, the calibration overhead of such approaches generally increases with system size. An alternative route is to identify the physical origins of crosstalk and suppress them at the hardware level.

Microwave crosstalk is a common form of signal crosstalk associated with the drive signals of superconducting qubits.
Prior works have modeled crosstalk mechanisms introduced by the sample package, including those arising from the printed circuit board (PCB)~\cite{Huang2021d} and from wire-bond-mediated coupling at the chip edges~\cite{Wenner2011a}.
However, a model that accounts for on-chip circuit elements and capable of quantitatively predicting the crosstalk at qubits remains to be developed.
On the other hand, microwave crosstalk has been characterized on various intermediate-scale quantum devices~\cite{Spring2021, Krinner2022, Karamlou2024, Kosen2024, Acharya2025} using qubits as a probe, but without particular focus on quantitative modeling.

In this work, we address this gap by systematically studying the physical mechanisms behind leading crosstalk effects in planar superconducting quantum devices.
Among all candidates, we identify two structures associated with dominating microwave crosstalk effects: a drive line routed in the proximity of other qubits, and a drive line crossing a qubit-qubit coupler using an air bridge crossover. 
We fabricate test devices featuring the two structures and measure the microwave crosstalk as a function of qubit frequency.
For each structure, we physically model the crosstalk and find agreement with the experiment.

Our results show that the crosstalk from a proximate drive line arises from two mechanisms: direct capacitive coupling between the drive line and the qubit island, and indirect coupling mediated by fields coupled into the sample package.
Consequently, accounting for the sample package geometry plays a crucial role in accurately modeling the microwave crosstalk.
In devices where a drive line crosses a qubit-qubit coupler with an air bridge crossover, microwave signals in the drive line can couple to the qubits connected by the coupler.
This mechanism coexists with the crosstalk from the proximate drive line, and the measured crosstalk can be accurately quantified by accounting for all contributions.

Finally, we discuss the implications of these findings for low-crosstalk device architectures.
We show that the stray field coupled into the package volume decays more slowly with distance compared to the capacitive crosstalk, making it difficult to scale planar devices while maintaining sufficiently low crosstalk-induced errors.
We outline several design strategies for suppressing microwave crosstalk, including the use of differential qubits and air-bridge crossovers with smaller footprints. 
In addition, advanced fabrication approaches such as through-silicon vias (TSVs)~\cite{Vahidpour2017, Yost2020, Mallek2021, Grigoras2022}, inter-chip metallic bumps~\cite{Rosenberg2017, Gold2021, Kosen2022, Norris2026}, and modular devices in which the elements are located on separate pieces of silicon~\cite{Field2024, Dalton2025} are expected to further reduce crosstalk by improving the confinement of the drive field.
The physical insights and analytical methods introduced in this work can be used to reduce microwave crosstalk in a wide range of superconducting quantum device applications from quantum computing~\cite{Wendin2017}, sensing~\cite{Lisenfeld2015, Besse2017, Wolski2020, Kristen2020}, quantum optics~\cite{Gu2017, Blais2021} to fundamental quantum science~\cite{Backes2021, Dixit2021, Braggio2025}.

\section{Crosstalk channels in planar superconducting devices} \label{sec:possible_crosstalk_channels}

Multiple microwave crosstalk channels can be present in superconducting quantum devices.
We outline several possible structures that can cause microwave crosstalk and discuss their relative magnitudes.

\begin{figure}[t]
    \includegraphics{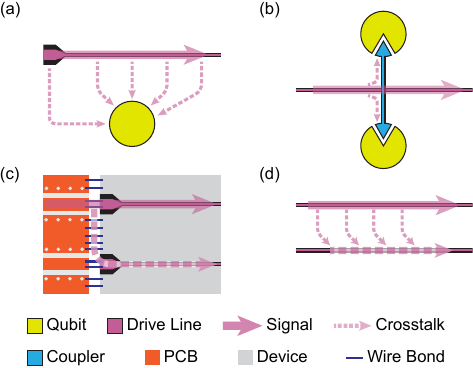}
    \caption{\label{fig:fig1_introdution}{\bf Microwave crosstalk channels.} 
    Schematic view of 
    {\bf (a)} a drive line (pink) routed in the proximity of a cross-coupled qubit (yellow), 
    {\bf (b)} a drive line crossing a qubit-qubit coupler (blue) with an air bridge crossover, 
    {\bf (c)} neighboring signal lines connected to the PCB (orange) with wire bonds (blue), 
    and {\bf (d)} signal lines running in parallel. 
    The drive (solid) and crosstalk (dashed) signals are indicated with transparent pink arrows.
    }
\end{figure}

Crosstalk arises when a drive line is routed in close proximity to another qubit.
In this case, the stray electromagnetic field of the drive signal can couple to the nearby qubit, as illustrated in Fig.~\ref{fig:fig1_introdution}a.
We refer to this effect as \emph{proximity-induced crosstalk}.
This effect is common in superconducting circuits and becomes increasingly pronounced in large-scale devices, where limited chip area constrains the separation between signal lines and qubits.
In Sec.~\ref{sec:proximity_induced_crosstalk}, we present the measurement and modeling of a test structure designed to study proximity-induced crosstalk.

Crosstalk can also be mediated by air bridge crossovers, which are commonly used in planar superconducting devices to route one signal line across another~\cite{Steffen2013, Chen2014j, Dunsworth2017a}.
At an air-bridge crossover, the upper signal line is routed across the lower line with a suspended metal bridge, introducing cross-capacitance between the two closely-spaced conductors.
In superconducting quantum processors, qubit-qubit couplers including coplanar waveguides (CPWs) are often added to enable two-qubit gates while increasing the physical separation between the qubits~\cite{Marques2021, Marxer2022, Krinner2022}.
When a drive line crosses such a coupler with an air bridge crossover, the drive signal can couple to the qubit-qubit coupler and subsequently drive the qubit connected to the coupler, as illustrated in Fig.~\ref{fig:fig1_introdution}b.
We refer to this mechanism as \emph{crossover-induced crosstalk}.
Despite being a second-order effect, crossover-induced crosstalk can be the most prominent crosstalk source on planar devices due to the considerable cross-capacitance and large capacitor pads used to realize significant qubit-qubit couplings.
In Sec.~\ref{sec:crossover_induced_crosstalk}, we discuss the measurement of a test structure where a drive line crosses a static qubit-qubit coupler and analytically model the crosstalk strength based on the design parameters.

In addition, crosstalk signals can propagate along the edge of the quantum device via the package-induced modes~\cite{Wenner2011a}.
In typical superconducting quantum computing systems, a gap is formed between the quantum device and its carrier PCB.
The gap supports traveling wave modes whose cutoff frequencies are increased by the wire bonds that galvanically connect the device and the PCB. 
With a sufficiently high wire-bond density, the lowest cutoff frequency can be raised above the operating frequencies of the signal lines.
Nevertheless, microwave drive signals can evanescently couple into these modes and propagate to adjacent lines, as illustrated in Fig.~\ref{fig:fig1_introdution}c.
We simulated our sample package and find this factor to be secondary compared to proximity-induced and crossover-induced crosstalk.
Details of this study are provided in Appendix~\ref{app:fem_simulation}1.

Also, microwave signals can in principle couple from one CPW to another nearby CPW, as illustrated in Fig.~\ref{fig:fig1_introdution}d.
In practice, this effect is strongly suppressed because CPWs provide efficient field confinement.
Moreover, in typical device layouts, the spacing between neighboring CPWs is more than ten times larger than the full CPW width, resulting in negligible inter-line coupling.
We confirm this expectation with numerical simulations based on the network parameters of CPWs in the quasi-static condition~\cite{Bazdar1994}, see Appendix~\ref{app:fem_simulation}2.

In summary, we identify proximity-induced crosstalk and crossover-induced crosstalk as the dominant microwave crosstalk in planar superconducting quantum devices.
Accordingly, the remainder of this work focuses on the experimental characterization and physical modeling of the two crosstalk channels.

\section{Quantifying microwave crosstalk} \label{sec:quantifying_microwave_crosstalk}

\begin{figure}[t!]
    \includegraphics{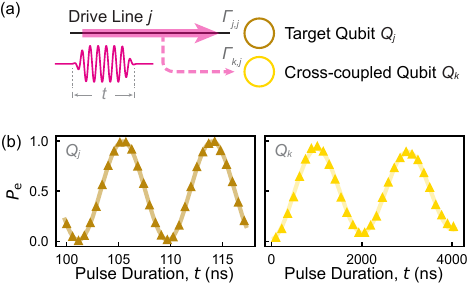}
    \caption{\label{fig:fig2_crosstalk_measurement}{\bf Crosstalk characterization.} 
    {\bf (a)} A drive line $j$ coupled to its target qubit $Q_j$ (brown) with coupling rate $\Gamma_{j,j}$ and to a cross-coupled qubit $Q_k$ (yellow) with coupling rate $\Gamma_{k,j}$.
    During the crosstalk characterization, a set of flat-top Gaussian filtered microwave pulses (pink) with a fixed amplitude and varying durations $t$ are applied from the drive line to excite the two qubits.
    {\bf (b)} Example of measured (triangles) excited-state population $P_{|e\rangle}$ of the target (brown) and cross-coupled (yellow) qubit as a function of drive pulse duration $t$. Solid lines of the same color show the fits to exponentially damped sinusoidal oscillations. 
    }
\end{figure}

We consider a drive line $j$ coupled to its target qubit $Q_j$ with coupling rate $\Gamma_{j,j}$ and to a cross-coupled qubit $Q_k$ with coupling rate $\Gamma_{k,j}$, see Fig.~\ref{fig:fig2_crosstalk_measurement}a.
We define cross-drive ratio $X_{k,j} \equiv \Gamma_{k,j} / \Gamma_{j,j}$~\cite{Spring2021, Kosen2024}, which is the fraction of drive power received by the cross-coupled qubit compared to the power received by the target qubit.

Experimentally, we extract $X_{k,j}$ by comparing the Rabi rates of the two qubits under the same set of drive pulses from drive line $j$.
We play a set of fixed-amplitude, varying-length flat-top resonant drive pulses, where the rising and falling edges are convolved with a Gaussian kernel of standard deviation $\sigma = \SI{5}{ns}$ to reduce spectral weight at other frequencies.
We record the excited-state population $P_\mathrm{e}$ as a function of the pulse duration $t$. 
The measured $P_\mathrm{e}(t)$ is fit to an exponentially damped sinusoidal oscillation to extract the Rabi rate of the target qubit $\Omega_{j,j}$ and of the cross-coupled qubit $\Omega_{k,j}$, see an example in Fig.~\ref{fig:fig2_crosstalk_measurement}b.
With the same drive pulses, the Rabi rate of the two qubits is proportional to the square root of the coupling rate, $\Omega \propto \sqrt{\Gamma}$~\cite{Blais2021}.
Hence, we calculate the experimentally-derived cross-drive ratio.
\begin{equation} \label{eq:crosstalk_ratio_definition}
X_{k,j} = \left(\frac{\Omega_{k,j}}{\Omega_{j,j}}\right)^2 \,.
\end{equation}
A lower cross-drive ratio implies weaker crosstalk, which is favorable for device operation.

\section{Proximity-induced crosstalk} \label{sec:proximity_induced_crosstalk}

\begin{figure*}[t!]
        \includegraphics{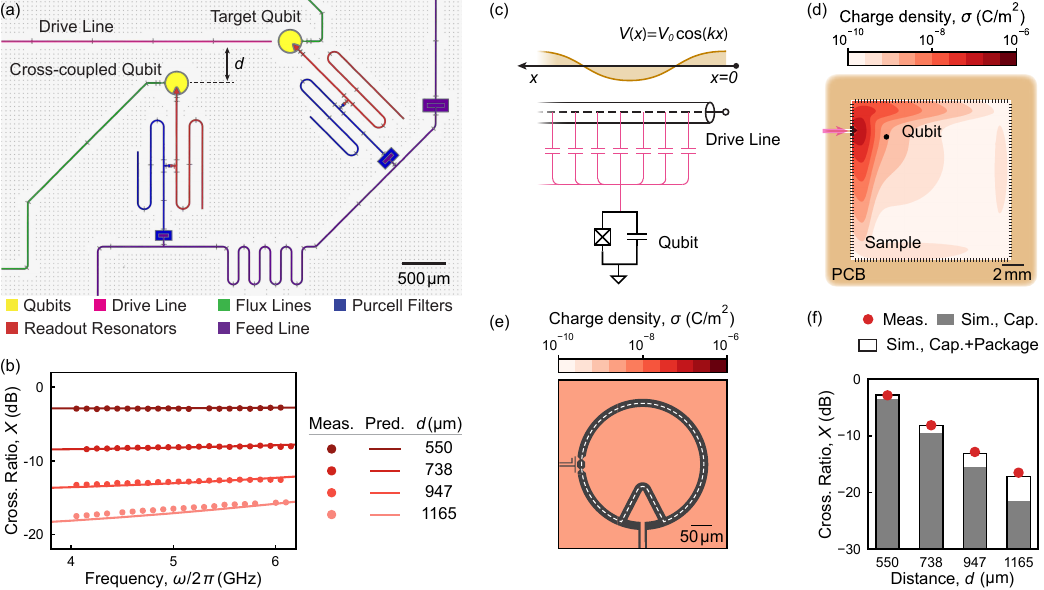}
    \caption{\label{fig:fig3_proximity_crosstalk}{\bf Characterizing and modeling the proximity-induced microwave crosstalk.} 
    {\bf (a)} False-colored micrograph of a test structure for the proximity-induced microwave crosstalk. $d$ denotes the distance between the drive line and the cross-coupled qubit. 
    {\bf (b)} Crosstalk drive ratio $X$ as a function of qubit frequency $\omega / 2\pi$ at different distances $d$. Circles show the measured crosstalk and the lines of the same color show the model.
    {\bf (c)} Schematic of driving the qubit via the direct capacitive coupling from the drive line. 
    The brown curve shows the distribution of the voltage amplitude as a function of length $x$ from the open end. The circuit highlighted in pink illustrates the distributed capacitance $c_\mathrm{dl}$ between each infinitesimal drive line segment and the transmon island.
    {\bf (d)} Simulated surface charge density over the ground plane when applying a $\SI{5}{GHz}$ drive with a unit power (pink arrow) to a shorted bonding pad corresponding to the drive line in panel (a). The sample is connected to the PCB (brown) with wire bonds (black). The vias connecting the top and bottom metalization layer of the PCB are not displayed.
    {\bf (e)} Simulated surface charge around the transmon island. The area where the metal is designed to be removed is shown in gray. The white dashed line encloses the area in which the surface charge is integrated to approximate the excited charge on the transmon island.
    {\bf (f)} Cross-drive ratio at $\SI{5}{GHz}$ obtained from the measurement (red circle), calculated considering both capacitive and package-mediated crosstalk (black wire frame), and calculated considering the capacitive crosstalk alone (filled bar).
    } 
\end{figure*}

We study the proximity-induced crosstalk using a test structure including a drive line and two frequency-tunable transmon qubits, see Fig.~\ref{fig:fig3_proximity_crosstalk}a.
The drive line is coupled to its target qubit, with a coupling rate characterized by the decay-limited lifetime $T_\mathrm{1,dl} = \SI{250}{\upmu s}$ at $\SI{5}{GHz}$.
A cross-coupled qubit is positioned at distance $d$ from the drive line, measured relative to the center of the transmon island. 
A dedicated flux line is added for each qubit to tune the frequency between 4 and \SI{6}{GHz}.
We characterize four test structures on a $\SI{14.3}{mm}\times \SI{14.3}{mm}$ superconducting quantum device.
The four test structures correspond to $d\in\{550, 738, 947, 1165\}\,\mathrm{\upmu m}$, with the cross-coupled qubit displaced along an axis orthogonal to the drive line.
At each characterized frequency, we measure the cross-drive ratio $X$ following the method described in Section~\ref{sec:quantifying_microwave_crosstalk}.
As expected, $X$ is suppressed at larger distances $d$, see Fig.~\ref{fig:fig3_proximity_crosstalk}b. 
We also observe an increase of crosstalk with frequency.

We consider two distinct mechanisms that contribute to the crosstalk observed in the test structure: direct capacitive coupling between the drive line and the qubit, and the indirect coupling from the spurious drive field inside the package volume.

First, to model the capacitive coupling, the drive line is treated as an open-ended semi-infinite transmission line.
The incident drive signal is reflected and forms a standing wave, resulting in the voltage distribution $V=V_0 \cos (kx)$, where $V_0$ is the maximum voltage amplitude and $x$ is the length from the open end. 
The wavenumber $k=\omega / v$ is determined by the frequency $\omega / 2\pi$ and the phase velocity $v$ in the CPW.
Each infinitesimal segment of the drive line is treated as an independent voltage source that capacitively drives the qubit, as illustrated in Fig.~\ref{fig:fig3_proximity_crosstalk}c. 
Using the electrostatic finite-element simulations described in Appendix~\ref{app:fem_simulation}3, we extract the capacitance per unit length $c_\mathrm{dl}(x)$ as a function of the position $x$. 
The effective coupling capacitance can be calculated with the integral
\begin{equation}
    C_\mathrm{dl,eff} = \int_{0}^{L} c_\mathrm{dl}(x) \cos (kx) \, \mathrm{d}x \, 
\end{equation}
over the length $L$ of the drive line. The cross-drive ratio $X$ from this mechanism is calculated with
\begin{equation}
    \sqrt{X_\mathrm{cap}} = \frac{C_\mathrm{dl,eff}}{C_\mathrm{dl,target}} \, ,
\end{equation}
where $C_\mathrm{dl,target}$ is the mutual capacitance between the drive line and its target qubit determined using electrostatic simulation.

Next, we model the crosstalk from the indirect coupling via the package volume. 
We perform radio-frequency finite-element simulations of a packaged device with just a bonding pad shorted to ground, see Appendix~\ref{app:fem_simulation}4.
The device is enclosed in a metal package and wire-bonded to a PCB, forming cavities above and below the sample, see details in Appendix~\ref{app:sample_and_setup}.
The corresponding resonance frequencies are usually referred to as the \emph{box modes}. 
We apply a unit excitation to the pad, and compute the surface charge distribution $\sigma$ on the ground plane, see Fig.~\ref{fig:fig3_proximity_crosstalk}d for the result at \SI{5}{GHz}.
Although this frequency is well below the lowest box-mode frequency of \SI{10.2}{GHz}, the drive signals still couple off-resonantly to the package volume and address the cross-coupled qubit. 

We integrate $\sigma$ within the midline of the gap around the transmon island, see Fig.~\ref{fig:fig3_proximity_crosstalk}e, to get $Q_\mathrm{c}$ which approximates the excited charge on the transmon island of the cross-coupled qubit. 
A similar simulation is performed for the target qubit driven by its designated drive line, yielding a reference charge $Q_\mathrm{t}$.

The simulations show that coupling via the package volume becomes significant only when a nonzero current flows through the signal line at the boundary of the sample, indicating that the coupling is inductive, see Appendix~\ref{app:fem_simulation}4.
We account for the current field amplitude at the bonding pad modulated by the standing wave in the drive line with a sinusoidal term $\sin(kL)$, where $L$ is the length of the drive line.
The package-mediated cross-drive ratio is estimated by scaling the excited charge accordingly,
\begin{equation} \label{eq:package_mediated_crosstalk}
    \sqrt{X_\mathrm{pac}} = \frac{Q_\mathrm{c}}{Q_\mathrm{t}} \sin (kL) \, .
\end{equation}

In the measured structure, both the stray capacitive coupling from the CPW and the package-mediated coupling contribute to the measured crosstalk. 
The two crosstalk drives superpose coherently, and we calculate the total proximity-induced crosstalk by summing their contributions,
\begin{equation}
\sqrt{X_\mathrm{prox}} = \sqrt{X_\mathrm{cap}} + \sqrt{X_\mathrm{pac}} \,.
\end{equation}
The calculated total crosstalk $X_\mathrm{prox}$ (solid lines) shows quantitative agreement with the experimental data, as shown in Fig.~\ref{fig:fig3_proximity_crosstalk}b. 
Importantly, both mechanisms are relevant to correctly explain the measured crosstalk: the crosstalk calculated from capacitive coupling alone systematically underestimates the observed crosstalk, with the relative discrepancy becoming more pronounced at larger separations $d$, see Fig.~\ref{fig:fig3_proximity_crosstalk}f.
Thus, at larger distances $d$, the package-mediated crosstalk becomes the dominant mechanism.
We will discuss the implications for device design in Section~\ref{sec:mitigation_strategies}.

\begin{figure}[t!]
    \includegraphics{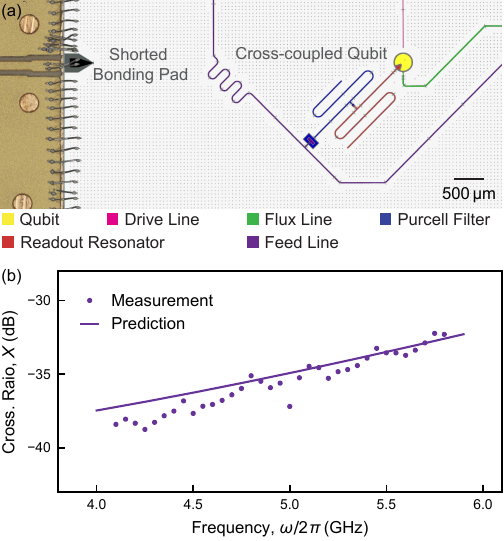}
    \caption{\label{fig:fig4_shorted_launcher_crosstalk}{\bf Driving the qubit via a shorted bonding pad.} 
    {\bf (a)} False-colored micrograph of the test structure where a cross-coupled qubit (yellow) is driven by a microwave signal applied to a shorted bonding pad (black).
    {\bf (b)} Measured (circles) and simulated (line) crosstalk drive ratio $X$ relative to the qubit's designated drive line as a function of the qubit frequency $\omega / 2\pi$.
    }
\end{figure}

To independently verify the model for the package-mediated crosstalk, we drive a qubit with a shorted bonding pad located at a distance of $\SI{5460}{\upmu m}$, see Fig.~\ref{fig:fig4_shorted_launcher_crosstalk}a.
The capacitive coupling in this case is negligible, and our model predicts the crosstalk to be dominated by the package-mediated coupling $X_\mathrm{pac}$.
Despite the low crosstalk level ranging from $-30$ to $\SI{-40}{dB}$, we observe quantitative agreement between the measured and simulated results, see Fig.~\ref{fig:fig4_shorted_launcher_crosstalk}b, providing further support for the proposed physical model. 

\section{Crossover-induced crosstalk} \label{sec:crossover_induced_crosstalk}

A simple and common qubit–qubit coupler is realized by connecting two qubits using a transmission line with capacitive coupler pads at the ends~\cite{Egger2019, Marques2021, Gold2021, Krinner2022, Karamlou2024, Norris2026}.
In planar two-dimensional qubit lattices, such couplers are inevitably crossed by microwave drive lines and can therefore introduce crosstalk. 

\begin{figure*}[t!]
    \includegraphics{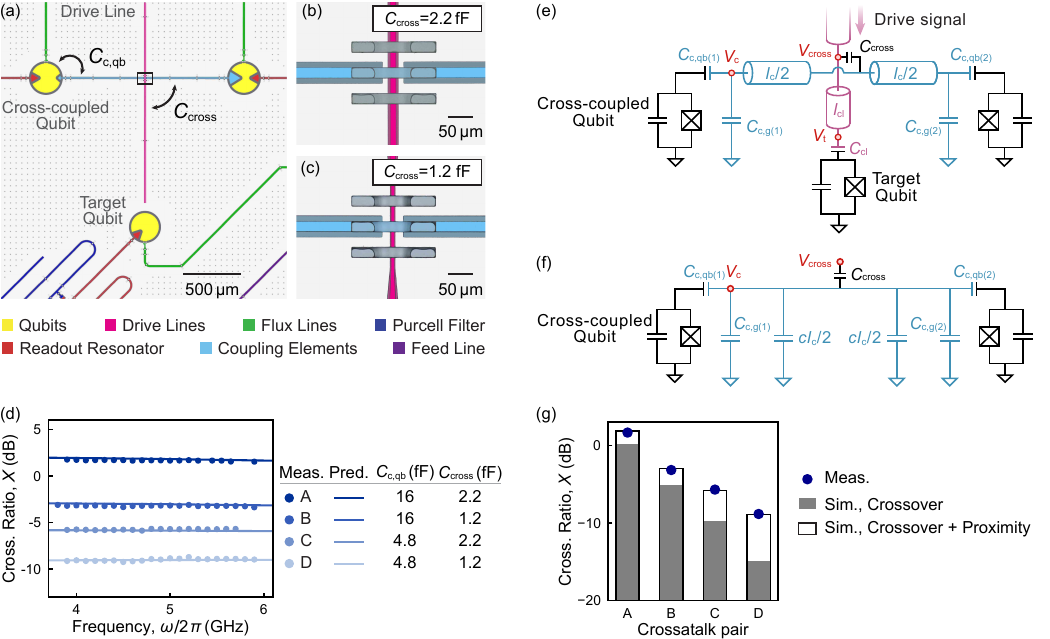}
    \caption{\label{fig:fig5_crossover_crosstalk}
    {\bf Characterizing and modeling the crossover-induced microwave crosstalk.} 
    {\bf (a)} False-colored micrograph of a test structure for the crossover-induced microwave crosstalk. Control parameters include the cross-capacitance $C_\mathrm{cross}$, which depends on the geometry of the air bridge crossover, and the coupler capacitance to qubit $C_\mathrm{c,qb}$ which depends on the coupler size.
    The black box marks the position of the air bridge crossover.
    In this experiment, we measured crossover designs with {\bf (b)} $C_\mathrm{cross}=\SI{2.2}{fF}$ and {\bf (c)} $C_\mathrm{cross}=\SI{1.1}{fF}$.
    {\bf (d)} Crosstalk drive ratio $X$ as a function of qubit frequency $\omega / 2\pi$ for different combinations of $C_\mathrm{c,qb}$ and $C_\mathrm{cross}$. Circles show the measured crosstalk, and the solid lines of the same color show the model. Labels of crosstalk pairs A to D correspond to the markers on the horizontal axis in panel (g).
    {\bf (e)} Circuit diagram for modeling the crossover-mediated crosstalk. The drive line (pink) is capacitively coupled to the qubit-qubit coupler (blue), and the coupler is capacitively coupled to the cross-coupled qubit. 
    {\bf (f)} Simplified circuit diagram of the qubit-qubit coupler and the cross-coupled qubits.
    {\bf (g)} Crosstalk ratio at $\SI{5}{GHz}$ obtained from the measurement (blue circles), and from the simulation including (black wire frames) and excluding (gray bars) the proximity-induced crosstalk.
    } 
\end{figure*}

To characterize this effect, we design test structures where both the center conductor and the ground planes of the coupler are routed across the drive line using air bridges, see Fig.~\ref{fig:fig5_crossover_crosstalk}a. 
This geometry introduces a cross-capacitance $C_\mathrm{cross}$ between the drive line and the coupler.
The coupler is connected to the qubits through capacitor pads with the coupling capacitance $C_\mathrm{c,qb}$. 
We predict $C_\mathrm{cross}$ with an electrostatic finite-element simulation that takes into account the height profile of the crossing air bridge, see Appendix~\ref{app:fem_simulation}5.
In this study, we fix the relative positions of the target qubit and the cross-coupled qubit, while varying the air-bridge crossover design to realize $C_\mathrm{cross} \in \{ 1.1, 2.2\} \, \mathrm{fF}$, see Fig.~\ref{fig:fig5_crossover_crosstalk}b,~c.
We also adjust the coupler pad size to obtain $C_\mathrm{c,qb} \in \{ 4.8, 16\} \, \mathrm{fF}$. 
This results in four combinations of $(C_\mathrm{cross}, C_\mathrm{c,qb})$, each corresponding to a different pair of drive line and cross-coupled qubit.
We characterize the microwave crosstalk for each pair as a function of frequency, measuring cross-drive ratios $X$ between $2$ and $-\SI{9}{dB}$, see Fig.~\ref{fig:fig5_crossover_crosstalk}d.
The crosstalk is reduced when decreasing the capacitances $C_\mathrm{cross}$ and $C_\mathrm{c,qb}$.

The measured crosstalk is dominated by the crosstalk signal reaching the cross-coupled qubits via the crossover and the coupler.
We model the drive line as a semi-infinite transmission line that couples to the target qubit with capacitance $C_\mathrm{cl}$ at the open end, and to the coupler with the cross capacitance $C_\mathrm{cross}$ at distance $l_\mathrm{cl}$ from the open end, see Fig.~\ref{fig:fig5_crossover_crosstalk}e. 
On each side of the crossing, the coupler CPW extends for a length $l_\mathrm{c}/2$ with capacitance $c$ to ground per unit length.
The coupler couples to the qubits via capacitor pads with coupling capacitance $C_{\mathrm{c,qb}(j)}, j\in\{1,2\}$, and each pad also has capacitance to ground $C_{\mathrm{c,g}(j)}$.

If the standing wave formed in the drive line has a voltage amplitude $V_\mathrm{t}$ at the open end, then the voltage amplitude at a crossover  is given by $V_\mathrm{cross} = V_\mathrm{t}\cos \left( kl_\mathrm{cl} \right)$.
To calculate the voltage amplitude $V_\mathrm{c}$ at the coupler capacitor pads, we consider a simplified circuit where the coupler CPW is replaced by its total capacitance to ground, and its series inductance is neglected. 
This treatment is valid in the quasi-lumped-element regime, $l_\mathrm{cl} \ll 2\pi/k$, and leads to the simplified circuit shown in Fig.~\ref{fig:fig5_crossover_crosstalk}f.
The voltage at the coupler pads is obtained from capacitive voltage division,
\begin{equation} \label{eq:voltage_division}
    V_\mathrm{c} = \frac{C_\mathrm{cross}}{\sum_j C_{\mathrm{c,g}(j)} + cl_\mathrm{c} + C_\mathrm{cross}} V_\mathrm{cross} \, .
\end{equation}
For a resonant drive applied through a capacitively coupled line, the qubit Rabi rate is given by~\cite{Krantz2019}
\begin{equation} \label{eq:drive_rabi_rate}
\Omega = \sqrt{\frac{\hbar \omega C_\Sigma}{2}} \frac{C_\mathrm{d}}{C_\Sigma} V \,
\end{equation}
where $\omega$ is qubit frequency, $V$ is the voltage amplitude of the drive signal, $C_\mathrm{d}$ is the coupling capacitance to the qubit, and $C_\Sigma$ is the total capacitance of the transmon island. 
In this experiment, all qubits in the test structures have a target $C_\Sigma=\SI{116}{fF}$, and each measurement is carried out with the target and cross-coupled qubit iteratively tuned to the same frequency.
Combining Eqs.~(\ref{eq:crosstalk_ratio_definition},~\ref{eq:voltage_division},~\ref{eq:drive_rabi_rate}), we can calculate the cross-drive ratio from the air bridge crossover
\begin{equation}
\sqrt{X_\mathrm{cross}} = \frac{C_{\mathrm{c,qb}(m)} C_\mathrm{cross}}{C_\mathrm{cl}\left(\sum_j C_{\mathrm{c,g}(j)}+cl_\mathrm{c} + C_\mathrm{cross}\right)} \cos \left(kl_\mathrm{cl}\right) \, .
\end{equation}
Here $m$ denotes the coupling capacitor to the cross-coupled qubit.

In the test structure, the drive line is inevitably routed near the cross-coupled qubit, and the proximity-induced crosstalk cannot be neglected.
We explicitly calculate the proximity-induced crosstalk contribution $X_\mathrm{prox}$ in addition to the crossover-mediated term.
We treat the system within the quasi-lumped-element regime and coherently sum the two contributions to obtain the total crosstalk,
\begin{equation}
    \sqrt{X_\mathrm{full}} = \sqrt{X_\mathrm{prox}} + \sqrt{X_\mathrm{cross}} \, .
\end{equation}
The calculated $X_\mathrm{full}$, shown with solid curves in Fig.~\ref{fig:fig5_crossover_crosstalk}d, quantitatively agrees with the experimental data.
We find that crossover-mediated crosstalk alone cannot account for the observed crosstalk magnitude, with the relative discrepancy becoming more pronounced as $X_\mathrm{cross}$ decreases, see Fig.~\ref{fig:fig5_crossover_crosstalk}g.
The crosstalk mediated by direct capacitance, via the package volume, and by the air bridge crossover should all be taken into account to correctly predict the total measured crosstalk.

\section{Suppressing microwave crosstalk} \label{sec:mitigation_strategies}

\begin{figure}[t!]
    \includegraphics{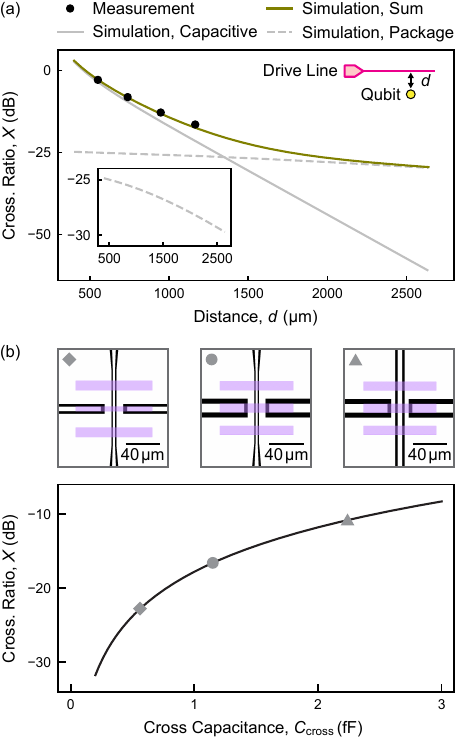}
    \caption{\label{fig:fig6_crosstalk_over_geometric_parameters}{\bf Reducing microwave crosstalk by adjusting geometric parameters.} 
    {\bf (a)} Proximity-induced crosstalk $X_\mathrm{prox}$ at $\SI{5}{GHz}$ as a function of distance $d$ between the drive line and the cross-coupled qubit.
    The olive line indicates the proximity-induced crosstalk, which is the combination of the capacitive (gray solid line) and package-mediated (gray dashed line) crosstalk.  
    The top right inset shows a schematic of a drive line (pink), a cross-coupled qubit (yellow) and the sweep parameter $d$.
    The bottom left inset presents the package-mediated crosstalk with a zoomed-in vertical axis.
    {\bf (b)} Crossover-induced crosstalk $X_\mathrm{cross}$ (solid line) at $\SI{5}{GHz}$ as a function of cross-capacitance $C_\mathrm{cross}$, assuming $C_\mathrm{c,qb}=\SI{7}{fF}$. 
    Markers highlight the 
    $X_\mathrm{cross}$ associated with crossover designs displayed in the insets, where black marks the area where the metal is removed and purple marks the air bridges. 
    } 
\end{figure}

As the amplitude of stray microwave fields generally decays with distance, a straightforward mitigation strategy for proximity-induced crosstalk is to maximize the separation between drive lines and cross-coupled qubits.
Based on the geometry shown in Fig.~\ref{fig:fig3_proximity_crosstalk}a, we simulate the crosstalk at \SI{5}{GHz} as a function of distance $d$ (olive line), see Fig.~\ref{fig:fig6_crosstalk_over_geometric_parameters}a.
We also present the contribution from the capacitive crosstalk (solid gray line) and from the package-mediated crosstalk (dashed gray line).
For distances $d < \SI{1000}{\upmu m}$, the capacitive contribution dominates the crosstalk.
As $d$ increases, this contribution is suppressed at a rate of \SI{-26}{dB/mm}, while the indirect coupling via the package volume, which decreases with distance only at rate \SI{-2.7}{dB/mm}, becomes dominant.
This slow spatial suppression implies that cross-coupled qubits with frequencies close to the working frequency of a drive line must be separated by large distances to reach a lower crosstalk-induced error, imposing more stringent layout constraints on planar superconducting devices.
In Appendix~\ref{app:layout_constraint_from_proximity_induced_crosstalk}, we further discuss how the interplay between the crosstalk drive ratio and qubit detuning determines the resulting single-qubit error and impacts the layout of low-crosstalk devices.

Alternative device architectures can be employed to overcome the limitations imposed by the proximity-induced crosstalk.
For example, using differential qubits~\cite{Place2021, Yanay2022, Pan2022, Bland2025} instead of grounded ones as in this study makes the cross-coupled qubits susceptible only to the gradient rather than to the amplitude of spurious field.
By aligning the dipole moment of differential qubits perpendicular to the gradient, the qubits can be made insensitive to proximity-induced crosstalk.
In addition, superconducting devices incorporating inter-chip metallic bumps~\cite{Rosenberg2017, Gold2021, Kosen2022, Norris2026}, TSVs \cite{Vahidpour2017, Yost2020, Mallek2021, Grigoras2022}, and with modular architectures~\cite{Field2024, Dalton2025} may improve the isolation of the drive signal and suppress the cross-driving of other qubits.

Crossover-induced crosstalk can be reduced by engineering the coupling elements along its propagation path.
One effective approach is to reduce the dimensions of the CPWs and air bridges at the crossover, as illustrated in Fig.~\ref{fig:fig6_crosstalk_over_geometric_parameters}b.
We simulate the cross-capacitance $C_\mathrm{cross}$ for three crossover designs: the measured large (triangle) and small crossover (circle) as well as a further miniaturized design (diamond) in which both the coupler CPW width and the air-bridge width are reduced by a factor of two relative to the small crossover.
These designs yield $C_\mathrm{cross} = \SI{2.2}{fF}$, $\SI{1.1}{fF}$, and $\SI{0.6}{fF}$, respectively.
Reducing $C_\mathrm{cross}$ from $\SI{2.2}{fF}$ to $\SI{0.6}{fF}$ is expected to result in an $\SI{11}{dB}$ reduction in the crosstalk contribution from the air bridge crossover, as shown in Fig.~\ref{fig:fig6_crosstalk_over_geometric_parameters}b.
Further reductions in the air bridge size may be achievable using advanced fabrication techniques such as gray scale e-beam lithography~\cite{Janzen2022b, VallesSanclemente2023}.

Microwave crosstalk can also be mitigated by engineering the length of the coplanar waveguide segments to create an destructive interference between the proximity-induced crosstalk and the crossover-induced crosstalk, thus canceling out the crosstalk at the operation frequency of the drive line.
In addition, in device architectures where neighboring qubits are arranged in far-detuned frequency bands, asymmetric qubit--qubit coupler pads, as shown in Fig.~\ref{fig:fig5_crossover_crosstalk}a, can be employed to suppress the crosstalk-induced error at the qubit whose frequency is closer to the operation frequency of the drive line.

\section{Conclusion and outlook}
In this work, we have systematically studied two structures that cause dominating microwave crosstalk in planar superconducting quantum devices.
We have fabricated test structures and characterized the crosstalk as a function of frequency.
Upon identifying the underlying physical mechanisms, we have developed models that quantitatively predict the crosstalk relying only on the device layout and sample package design.
The model also correctly explains the strongest crosstalk instances in a 17-qubit quantum processor, see Appendix~\ref{app:crosstalk_on_s17}.
Our analysis reveals that a drive signal can address a nearby qubit either through stray capacitance from the drive line or via indirect coupling through the package volume.
The latter contribution is suppressed less with increasing distance, thereby imposing more constraints on qubit placement and signal-line routing.
Based on these insights, we have discussed design strategies for suppressing the identified crosstalk mechanisms.

This work establishes a physically grounded and quantitative framework for understanding microwave crosstalk in planar superconducting devices. 
The resulting models provide practical guidance for the design of quantum devices with reduced microwave crosstalk.
Building on these models, future studies can extend the analysis to weaker and higher-order crosstalk channels, with the goal of developing a more complete understanding of microwave crosstalk in superconducting circuits.
The analytical framework developed here are not limited to superconducting quantum processors, but are also applicable to a broader class of superconducting platforms which require good isolation between components and selectivity of control signals.

\section*{Acknowledgments}

The authors thank Ilya Besedin, Michael Kerschbaum, François Swiadek, Dominik Haegi, Nicolas Thill, and Moritz Bürgi for contributing to the crosstalk characterization on 17-qubit devices. The authors acknowledge financial support by the Intelligence Advanced Research Projects Activity (IARPA) and the Army Research Office, under the Entangled Logical Qubits program and Cooperative Agreement Number W911NF-23-2-0212, by the Baugarten Foundation, the ETH Zurich Foundation, and by ETH Zurich.
The views and conclusions contained in this document are those of the authors and should not be interpreted as representing the official policies, either expressed or implied, of IARPA, the Army Research Office, or the U.S. Government. The U.S. Government is authorized to reproduce and distribute reprints for Government purposes notwithstanding any copyright notation herein.

\section*{Author contribution}
Y.\,S. and M.\,P. planned the experiments. 
Y.\,S. designed the devices with the input from M.\,P., K.\,D., and A.\,W..
D.\,H., K.\,D., F.\,H., F.\,W., and M.\,B.\,P. fabricated the devices.
Y.\,S. prepared the measurement setup and acquired the data.
Y.\,S. and M.\,P. performed the analysis.
M.\,P. and A.\,W. supervised the project.
Y.\,S. and M.\,P. wrote the manuscript with inputs from all authors.

\appendix
\section*{Appendices}
\renewcommand{\thesubsection}{\thesection\arabic{subsection}}

\section{Simulation of microwave crosstalk} \label{app:fem_simulation}

\subsection{Crosstalk mediated by the chip edge mode} \label{app:chip_edge_crosstalk}

\begin{figure}[b!]
    \includegraphics{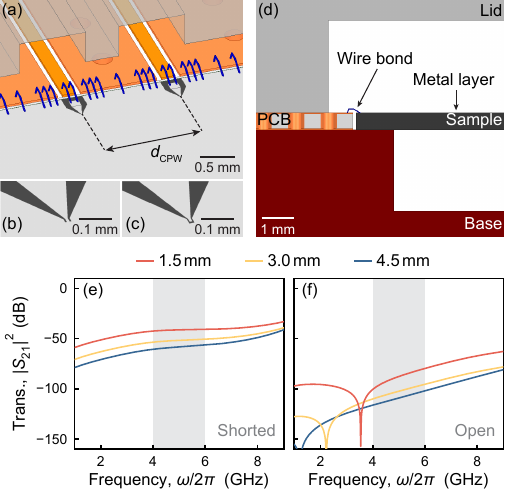}
    \caption{\label{fig:fig7_launcher_crosstalk_sim}
    {\bf Simulation of microwave crosstalk mediated along the chip edge.} 
    {\bf (a)} In the finite-element simulation model, the sample (light gray) is connected to the PCB (orange) with wire bonds (blue).
    A lid (gray, transparent) covers the PCB and has cutouts for the signal lines.
    Two bonding pads, separated by distance $d_\mathrm{CPW}$, are either both {\bf (b)} shorted to ground or {\bf (c)} left open in the simulation.
    {\bf (d)} Cross-section view of the model, including the sample package formed by the aluminum lid (gray) and the copper base (brown).
    The transmission is simulated between the two CPW trances terminated by {\bf (e)} shorted and {\bf (f)} open bonding pads with separations $d_\mathrm{CPW}=\SI{1.5}{mm}$ (red), $\SI{3.0}{mm}$ (yellow), and $\SI{4.5}{mm}$ (blue).
    The shaded area highlights the frequency band in which the crosstalk is experimentally studied in this work.
    } 
\end{figure}

We study the microwave crosstalk mediated at the chip edge between two signal lines with a separation $d_\mathrm{CPW}$, see Fig.~\ref{fig:fig7_launcher_crosstalk_sim}a, which is a candidate source of microwave crosstalk but significantly weaker than the ones presented in the main text.
The metal layer of the sample forms a distributed capacitance to the PCB and the copper base, which creates a traveling wave mode at the edge of the sample together with the self-inductance of the wire bonds. The mode has a finite cutoff frequency which is typically above the frequencies of interest for qubit driving. It therefore allows crosstalk via evanescent waves, with a coupling strength decreasing exponentially with distance~\cite{Wenner2011a}. Details of the sample package are presented in Appendix~\ref{app:sample_and_setup}.

The coupling to the chip edge mode is inductive, and the amplitude of the crosstalk signal is proportional to the current amplitude at the chip edge.
Accordingly, we consider two extreme cases in the simulation: the bonding pads at the end of the signal lines are either both shorted to ground, see Fig~\ref{fig:fig7_launcher_crosstalk_sim}b, which leads to a current anti-node at the wire bonds, or both left open, see Fig~\ref{fig:fig7_launcher_crosstalk_sim}c, which creates a current node at the wire bonds.
We expect the crosstalk to reach its maximum between the shorted bonding pads, and minimum between the open bonding pads.
In realistic samples, the crosstalk mediated by the chip edge mode is expected to be in between the two extreme cases.

We account for the geometry of the sample package including the aluminum lid, the PCB, the copper base, the sample, and the wire bonds, see the cross-section view in Fig.~\ref{fig:fig7_launcher_crosstalk_sim}d. 
The metal layer of the sample is modeled as a perfectly conducting sheet with zero thickness, and all other metal parts are modeled as bulky perfect conductors.
The sample substrate, made of high-resistivity silicon, is modeled as a piece of $\SI{520}{\upmu m}$-thick dielectric with relative permittivity $\varepsilon_\mathrm{r}=11.45$. The PCB substrate, made of ceramic-filled laminates RO3210, is modeled as a dielectric with $\varepsilon_\mathrm{r}=10.2$.
We simulate the transmitted power $|S_{21}|^2$ from one signal line to the other between $1\sim\SI{9}{GHz}$, covering the operation frequency for most superconducting quantum devices.

We choose separations $d_\mathrm{CPW}\in \{1.5, \, 3.0, \, 4.5\}\, \mathrm{mm}$ in the simulation. 
With the shorted bonding pads, the crosstalk transmission increases with frequency $\omega / 2\pi$ and decreases with the separation $d_\mathrm{CPW}$, see Fig.~\ref{fig:fig7_launcher_crosstalk_sim}e.
With the open bonding pads, the transmitted power is more than three orders of magnitude lower, see Fig.~\ref{fig:fig7_launcher_crosstalk_sim}f, in agreement with our expectation that this coupling is inductive. 
On the experimental device, $d_\mathrm{CPW}=\SI{3.0}{mm}$, and the crosstalk between neighboring signal lines is upper-bounded by that between two shorted lines, which remains below $\SI{-50}{dB}$ in the $4\sim 6\,\mathrm{GHz}$ qubit frequency range.
Thus, we conclude that crosstalk via the chip edge mode is expected to be significantly weaker than the crosstalk mechanisms studied in the main text.

\subsection{Signal crosstalk between parallel CPWs}

In this appendix, we discuss another candidate of microwave crosstalk, the cross-transmission between proximate signal lines on the device. 
In the experimental device, signals in the CPWs predominantly propagate via the quasi-TEM mode~\cite{Simons2001}. 
Other modes, such as the slotline modes, are suppressed by connecting the ground on the two sides of a CPW with air bridges.
The quasi-TEM mode confines the signal to the proximity of the CPW, efficiently suppressing the microwave crosstalk between two CPWs when the distance between them is increased.

\begin{figure}[b!]
    \includegraphics{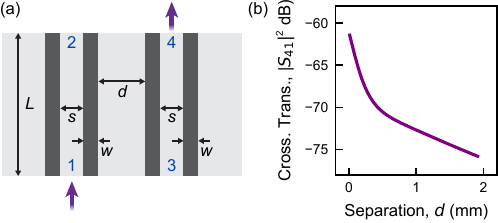}
    \caption{\label{fig:fig8_cpw_crosstalk}
    {\bf Crosstalk between parallel CPWs.}
    {\bf (a)} Geometric parameters for simulation. 
    Each CPW is defined with its center conductor width $s=\SI{10}{\upmu m}$ and gap width $w=\SI{5.5}{\upmu m}$.
    The two CPWs are separated with a ground plane with variable width $d$ and the parallel segment extends for length $L=\SI{5}{mm}$.
    The crosstalk transmission is extracted from port 1 to port 4 (purple arrows).
    {\bf (b)} Crosstalk transmission $|S_\mathrm{41}|^2$ as a function of $d$ calculated from a numerical model while fixing the signal frequency at $\SI{5}{GHz}$.
    } 
\end{figure}

We consider two CPWs running in parallel with a separation $d$, each characterized by the center conductor width $s=\SI{10}{\upmu m}$, the gap width $w=\SI{5.5}{\upmu m}$, and length $L = \SI{5}{mm}$, see Fig.~\ref{fig:fig8_cpw_crosstalk}a.
We simulate the transmitted power $|S_{41}|^2$ from 
port 1 on one end of a CPW to port 4 which is on the opposite end of the neighboring CPW.
With typical separations $d$ on realistic devices, this crosstalk is too low to be efficiently simulated with finite-element modeling tools, so we instead numerically calculate the crosstalk transmission from a circuit model where the network parameter of parallel CPW segments are expressed analytically in the quasi-static condition~\cite{Bazdar1994}.
$|S_{41}|^2$ is predicted to be $-\SI{65}{dB}$ when the two drive lines are separated by $d=\SI{0.1}{mm}$ and drops below $-\SI{72}{dB}$ at $d>\SI{1}{mm}$, see Fig.~\ref{fig:fig8_cpw_crosstalk}b.
Keeping the separation above \SI{0.1}{mm} is possible on many intermediate-scale quantum devices.
For instance, the smallest distance between signal lines is \SI{0.25}{mm} on the 17-qubit quantum processor described in Ref.~\cite{Krinner2022}.
Thus, we conclude that the microwave crosstalk between on-chip signal lines is negligible compared with the mechanisms discussed in the main text.

\subsection{Direct capacitance between a drive line and a transmon island} \label{app:distributed_drive_line_capacitance}

A drive line could cross-couple to a qubit via direct capacitance.
Following the discussion in Section~\ref{sec:proximity_induced_crosstalk}, we split the drive line into infinitesimal segments and treat each segment as an individual voltage source that capacitively drives the qubit.
As the drive line is an open-ended transmission line, the drive signal will be reflected and form a standing wave, which modulates the voltage amplitude along the line.
This modifies the crosstalk drive rate, and requires us to include the spatial dependence of the direct capacitance into the model.
First, with an electrostatic finite-element simulation, we model a $\SI{10}{mm} \times \SI{10}{mm}$ sample where the transmon island is located at the center.
We treat the sample metal layer as a \SI{120}{nm}-thick perfect conductor, and the substrate as a dielectric as described in Appendix~\ref{app:fem_simulation}1.
We include a vacuum gap of \SI{2}{mm} above and below the sample and introduce Dirichlet boundaries on the exterior of the modeled area.
The dimensions of the model is chosen such that further increasing the simulated domain would negligibly increase the capacitance of the transmons island.
We assign a unit voltage excitation to the transmon island and simulate the surface charge on the upper and lower surface of the ground plane.
The sum of the charge, $c(\bm{r})$, represents the capacitance between the transmon island and a piece of ground plane with unit area at position $\bm{r}$.

\begin{figure}[t!]
    \includegraphics{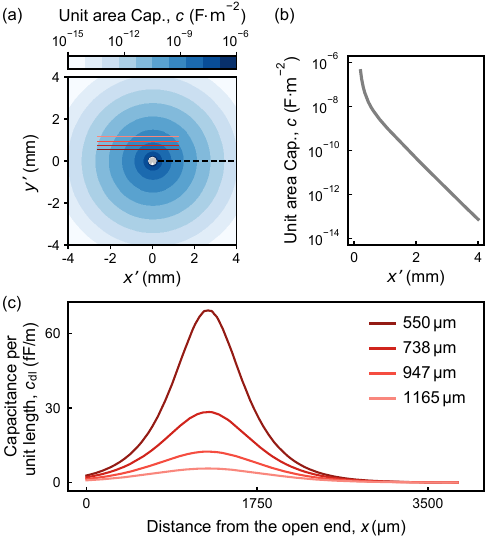}
    \caption{\label{fig:fig9_distributed_capacitance_to_qubit}
    {\bf Simulated direct capacitance to the qubit.} 
    {\bf (a)} Capacitance $c$ between the transmon island (light gray) and unit area of ground plane as a function of coordinate $\bm{r}=(x', y')$ relative to the center of the qubit.
    Horizontal solid lines marks the relative position of the drive lines studied in the test sample for the proximity-induced microwave crosstalk.
    {\bf (b)} Capacitance per unit area $c$ along the $+x'$ axis, indicated by the black dashed line in panel (a).
    {\bf (c)} Capacitance per unit length of drive line to the transmon island, as a function of distance $x$ from the open end. The colors distinguishes the four tested distances $d$ between the cross-coupled qubit and the drive line, whose positions are highlighted in panel (a).
    } 
\end{figure}

Because the transmon island is approximately circular, $c(\bm{r})$ is highly rotationally symmetric about the island center. 
For distances $|\bm{r}|>\SI{0.5}{mm}$, $c(\bm{r})$ decays exponentially with distance, as shown in Fig.~\ref{fig:fig9_distributed_capacitance_to_qubit}a..
This rapid decay implies that direct capacitance can be effectively reduced by placing other drive lines farther from the qubit.
Taking the capacitance along the $+x$ axis as an example, relative to its value at $x=\SI{0.5}{mm}$, $c$ is reduced by factors of 9, 267, and 6680 at $x=1,\, 2,\, 3\, \mathrm{mm}$, as shown in Fig.~\ref{fig:fig9_distributed_capacitance_to_qubit}b.

We note that while we primarily relied on FEM simulations for the results of this appendix, certain settings (e.g. evaluation of many different qubit shapes) may benefit from a faster method to calculate the mutual capacitance per unit area. An exact analytic solution exists if we ignore the gaps in the metal film, assume a circular transmon island, and re-frame the problem as finding the electrostatic \emph{Dirichlet-to-Neumann kernel} (the charge density induced on a perfectly conducting boundary with a potential given by the delta distribution) for an infinite slab (the vacuum volume above the metal layer) or a pair of stacked slabs (the substrate and the vacuum volume below the chip). For the single slab, the result takes the form
\[
 c(\bm{r}) = \frac{\varepsilon \pi}{h^3}\sum_{n=1}^{\infty}
 n^2 K_0(n\pi |\bm{r}|/h)
\]
with $\varepsilon$ the slab permittivity, $h$ its height and $K_0$ a modified Bessel function of the second kind. Due to the exponential fall-off of $K_0$, this result also makes it clear that for $|\bm{r}|$ exceeding $h$, the term $n=1$ dominates and the capacitance decays exponentially with distance. For the case of a stack of multiple slabs, a similar result holds but the coefficients $n$ in the argument of $K_0$ are replaced by the eigenvalues of a 1D boundary value problem associated with the structure of the stack.

Around a coplanar waveguide, the charge is redistributed with respect to the solution on the continuous ground plane as the metal is removed at the gaps. 
We treat this as a perturbation that does not affect the overall charge distribution, and assume that the charge originally on the removed metal is evenly redistributed onto the center conductor and the ground.
The direct capacitance per unit length of drive line at coordinate $\bm{r}$ is given by $(w+s)c(\bm{r})$, where $s$ and $w$ are the widths of the center conductor and the gap.
With this, we calculate the capacitance $c_\mathrm{dl}$ per unit length of drive line to the transmon island as a function of distance $x$ from the open end, see Fig.~\ref{fig:fig9_distributed_capacitance_to_qubit}c.
The capacitances $c_\mathrm{dl}(x)$ calculated for the four test structures all reach their maximum at the same position $x$ which corresponds to the point with minimum distance from the transmon island.

\subsection{Cross-driving mediated over the sample package}

\begin{figure}[t!]
    \includegraphics{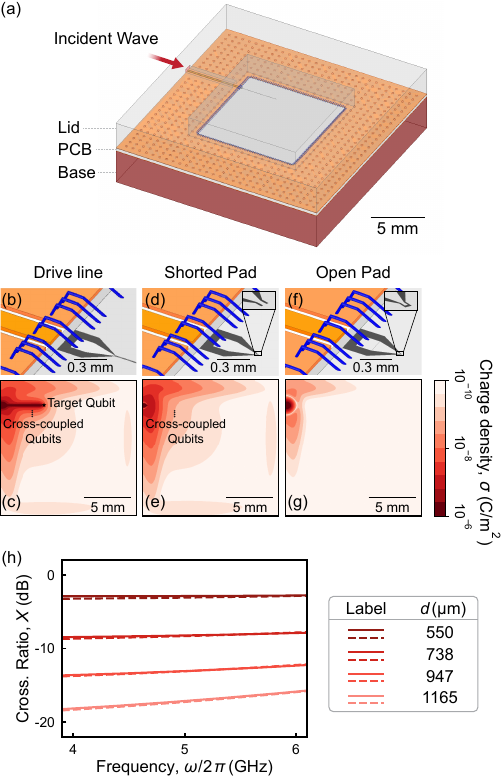}
    \caption{\label{fig:fig10_crosstalk_via_package_mode}
    {\bf Crosstalk via the sample package.}
    {\bf (a)} The sample package and the sample are modeled in a finite-element simulation. 
    An incident wave (red arrow) with unit power is applied to a CPW trace, which is galvanically connected to a bonding pad on the sample with wire bonds.
    The bonding pad is either {\bf (b)} connected to a drive line, or {\bf (d)} shorted to the ground, or {\bf (f)} left open.
    The insets show enlarged views of the ends of the bonding pads.
    {\bf (c,e,g)} Simulated charge density on the ground plane at \SI{5}{GHz} for the three corresponding bonding pad terminations: connected to a drive line, shorted to ground, and left open, respectively.
    Black markers indicate the positions of the target qubit (star) and cross-coupled qubits (circles) in the proximity-induced-crosstalk test sample.
    {\bf (h)} Cross-drive ratio $X$ predicted by separately modeling the capacitive and package-mediated crosstalk (solid lines) and extracted combined from the simulation including a drive line (dashed lines). The color schemes distinguishes the different distances $d$ between the drive line and the cross-coupled qubit.
    } 
    \vspace{-8pt}
\end{figure}

Microwave crosstalk could appear when the drive signal couples into the package volume and addresses the cross-coupled qubits.
We set up a simulation at RF frequencies following the sample package and material properties described in Appendix~\ref{app:fem_simulation}1. 
We consider one signal line formed by wire-bonding a PCB CPW trace to a bonding pad, see Fig.~\ref{fig:fig10_crosstalk_via_package_mode}a.

In the first simulation, we connect the bonding pad to a drive line of the proximity-induced crosstalk test sample, see Fig.~\ref{fig:fig10_crosstalk_via_package_mode}b.
A drive signal with unit power is applied to the signal line and the charge density over the ground plane of the sample is extracted from the simulations, see Fig~\ref{fig:fig10_crosstalk_via_package_mode}c for the solution with a \SI{5}{GHz} drive signal.
As expected, the field is the strongest around the drive line.
Despite the fact that the drive frequency is significantly lower than the lowest resonance frequency of the package $\omega_\mathrm{box, 0} / 2\pi = \SI{10.2}{GHz}$, the drive signal can still penetrate towards the chip center with a suppression over the distance.
We can derive the cross-driving from this model by comparing the excited charge on the cross-coupled and the target qubit, where the contribution from the direct capacitance, as discussed in Appendix~\ref{app:fem_simulation}3, and from the indirect coupling via the sample package, are both included.

In the second simulation, we short the bonding pad to ground, see Fig.~\ref{fig:fig10_crosstalk_via_package_mode}d, to independently model the indirect cross-driving via the sample package.
Despite removing the drive line, the feature of the drive field penetrating towards the chip center persists, see Fig.~\ref{fig:fig10_crosstalk_via_package_mode}e. 
This shows that the coupling via the sample package is a mechanism that co-exists with crosstalk mechanisms defined by the circuits in the sample.

In the third simulation, we leave the bonding pad open, see Fig.~\ref{fig:fig10_crosstalk_via_package_mode}f. 
We observe that the drive field towards the chip center is largely suppressed, see Fig.~\ref{fig:fig10_crosstalk_via_package_mode}g, which shows that coupling into the sample package is predominantly inductive.

The system formed by the sample package and the ground plane is linear, so we take the excited charge at the cross-coupled qubits mediated by the sample package to be proportional to the current at the edge of the chip.
We calculate $Q_\mathrm{c}$ based on the charge distribution simulated with the shorted bonding pad, multiplying with the scaling factor $\sin (kL)$ to account for the current amplitude at the bonding pad introduced by the standing wave, and arrive at Eq.~(\ref{eq:package_mediated_crosstalk}) for evaluating the package-mediated crosstalk $X_\mathrm{pac}$.

Notably, we find that calculation of the crosstalk by summing the capacitive contribution described in Appendix~\ref{app:fem_simulation}3 with the package-mediated contribution simulated with the shorted bonding pad gives a result almost identical to the simulation that includes the full drive line, see Fig.~\ref{fig:fig10_crosstalk_via_package_mode}h. This gives us confidence in separating the overall crosstalk into the two contributions.

\subsection{Cross-capacitance at the air bridge crossover}

\begin{figure}[b!]
    \includegraphics{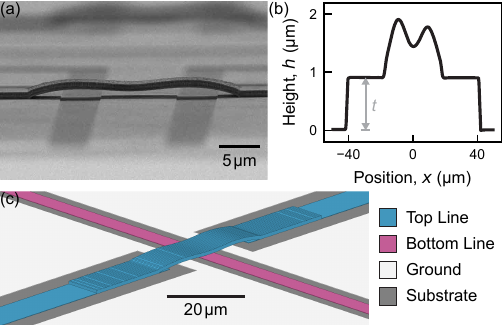}
    \caption{\label{fig:fig11_ab_height_profile}
    {\bf Cross-capacitance at the crossover.}
    {\bf (a)} Scanning electron microscope image of an air bridge crossing a coplanar waveguide.
    {\bf (b)} Measured height profile along center of the upper surface of a crossover air bridge. 
    Height $h=0$ corresponds to the upper surface of the base metal layer. 
    Thickness $t$ of the air bridge metal is marked with arrows in light gray.
    {\bf (c)} Crossover structure is modeled in an electrostatic simulation. The cross-capacitance $C_\mathrm{cross}$ is extracted between the top line (blue) crossing over the bottom line (pink).
    } 
\end{figure}

Air bridges are usually fabricated on planar superconducting devices to connect ground planes across a CPW, see Fig.~\ref{fig:fig11_ab_height_profile}a, or realize crossings between signal lines.
In the second case, the two signal lines will acquire a cross-capacitance due to the proximity between their CPW traces and between the crossing air bridge and the bottom line.
Extracting the cross-capacitance $C_\mathrm{cross}$ is necessary for calculating the crosstalk mediated by the air bridge crossover.
To obtain the height profile of an air bridge, we image the top surface of the air bridges with a laser scanning microscope.
Along the center line of the bridge, we find a step increase corresponding to the metal thickness $t$, and then a profile corresponding to the suspended metal segment, see Fig.~\ref{fig:fig11_ab_height_profile}b.

For the extracted geometry, we set up an electrostatic simulation with a qubit-qubit coupler (blue) crossing a drive line (pink), see Fig.~\ref{fig:fig11_ab_height_profile}c.
The CPW sections in the simulation are sufficiently long ($600\,\mathrm{\upmu m}$) such that increasing the dimensions further has negligible impact to the extracted $C_\mathrm{cross}$.
We simulate $C_\mathrm{cross}=\SI{2.2}{fF}$ from the larger air bridge crossover geometry in the test samples for crossover-induced crosstalk.
We infer that approximately $70\%$ of the cross-capacitance is contributed by the capacitor formed between the air bridge and the drive line, while the remaining $30\%$ is due to the proximity of the CPW part of the two lines.

\section{Quantum device and experimental setup} \label{app:sample_and_setup}

We perform the experiments on superconducting quantum devices with dimensions of $\SI{14.3}{mm}\times\SI{14.3}{mm}$. 
The transmon islands, coplanar waveguides, and couplers are patterned with photolithography and reactive-ion etching into a 120-nm niobium thin film sputtered onto a high-resistivity silicon substrate. Aluminium-titanium-aluminium trilayer airbridges connect the otherwise-separated ground planes. 
Using electron-beam lithography (EBL) and shadow evaporation, we fabricate SQUID loops formed by Manhattan-style Josephson junctions~\cite{Potts2001, Costache2012, ColaoZanuz2025} connecting the transmon islands and the ground. These define the frequency-tunable qubit modes.

\begin{figure}[b!]
    \includegraphics{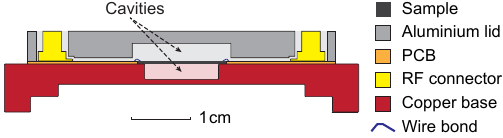}
    \caption{\label{fig:fig12_chip_crosssection}{\bf Cross-section layout of the sample packaging.} The sample (black) is attached to the copper base (red) and wire-bonded to the printed circuit board (orange). An aluminum lid (light gray) shields the sample from the top. Cavities are added below and above the sample. Signals form RF connectors (yellow) are routed to the sample with signal lines on the PCB and the wire bonds (blue, barely visible as the element is small). 
    } 
\end{figure}

\begin{figure}[t!]
    \includegraphics{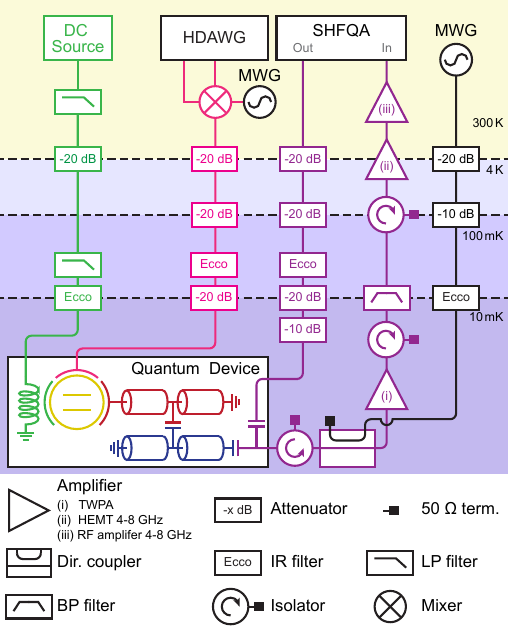}
    \caption{\label{fig:fig13_cabling_diagram}
    {\bf Schematic diagram of the experimental setup.} 
    Flux lines (green), drive lines (pink), and readout lines (purple) transmit the signal between the control electronics and the device, which holds transmon qubits (yellow), readout resonators (red), and Purcell filters (blue).  
    The background colors indicate the temperature stages of the cryostat.
    } 
\end{figure}

Each quantum device under characterization is attached to a copper base and connected to coaxial control lines by a carrier PCB, see Fig.~\ref{fig:fig12_chip_crosssection}.
Circular aluminum wire bonds with a diameter of $\SI{25}{\upmu m}$ are added to connect the sample and the PCB over the \SI{0.1}{mm} gap.
An aluminum lid is installed on top of the sample to provide mechanical protection and shield the sample from the environmental electromagnetic field.
Cavities are introduced above and below the sample, with fundamental resonance frequencies $\omega_\mathrm{0,above} / 2\pi = \SI{13.3}{GHz}$ and $\omega_\mathrm{0,below} / 2\pi = \SI{10.2}{GHz}$ expected from simulation.
These frequencies are significantly higher than the $4\sim 6\, \mathrm{GHz}$ qubit frequency band and $6.5\sim 7.5\, \mathrm{GHz}$ readout resonator frequency band in the experimental device, which suppresses the qubit and resonator decoherence from the package resonance modes.

The quantum devices are installed at the $\SI{10}{mK}$ stage of a dilution refrigerator with a standard wiring and shielding configuration~\cite{Krinner2019}.
We connect the room-temperature electronics to the device with signal lines comprising microwave components and coaxial cables, see Fig.~\ref{fig:fig13_cabling_diagram}.
The DC source provides a bias current for each qubit to set their operation frequency.
The arbitrary waveform generator (HDAWG) generates baseband signals with a \SI{2.0}{GSa/s} sampling rate.
The baseband singals are up-converted to microwave drive signals with IQ-modulation incorporating a microwave generator (MWG) and a mixer.
We use a readout instrument (SHFQA) to generate probe signals to the readout resonators. 
The resonator response passes through an amplification chain consisting of a wide-bandwidth near-quantum limited traveling-wave parametric amplifier (TWPA)~\cite{Macklin2015}, a high-electron-mobility transistor (HEMT) and a room-temperature amplifier before acquired by the same readout instrument.
The acquired signal is averaged and compared with the reference signals from preparing the qubits in the ground and excited state to reconstruct the population of the two qubit states.

\section{Single-qubit error introduced by proximity-induced crosstalk}
\label{app:layout_constraint_from_proximity_induced_crosstalk}

Microwave crosstalk signals can cause single-qubit errors by coherently rotating the cross-coupled qubits.
Assuming no other pulses played at the same time, the introduced error is associated with the cross-drive ratio $X$ and the detuning $\Delta_\mathrm{dq}$ between the drive line's operation frequency and the qubit frequency.
We quantify this effect by studying the error introduced on a cross-coupled transmon qubit when a X180 rotation pulse is applied via the drive line.
We assume a qubit anharmonicity of $\alpha/2\pi = -\SI{165}{MHz}$, and a Gaussian drive pulse envelope with standard deviation $\sigma=\SI{9.6}{ns}$ truncated at $\pm 2.5\sigma$, consistent with the experimental setup.

\begin{figure}[t!]
    \includegraphics{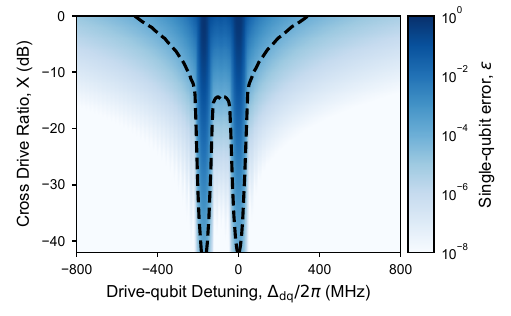}
    \caption{\label{fig:fig14_numerical_qutrit_error}
    {\bf Crosstalk-induced single-qubit error.} 
    We assume that a Gaussian-envelope drive pulse with $\sigma=\SI{9.6}{ns}$ truncated at $\pm 2.5\sigma$ is applied to the target qubit, and simulate single-qubit errors $\varepsilon$ at an idling cross-coupled qubit as a function of detuning $\Delta_\mathrm{dq}$ and cross-drive ratio $X$, averaged over the six cardinal initial states. Black dashed line marks the contour $\varepsilon=10^{-4}$.
    In the simulation, we set the anharmonicity of the cross-coupled qubit to be $\alpha/2\pi = -\SI{165}{MHz}$.
    } 
\end{figure}

We simulate the evolution of the cross-coupled qubit with the QuTiP master equation solver~\cite{Johansson2012a}, and average the introduced single-qubit errors over the six cardinal states $|0/1\rangle, \, |\pm\rangle, \, |\pm i\rangle$. 
Prominent errors occur when the drive is resonant with the g-e and e-f transitions of the qubit, see Fig.~\ref{fig:fig14_numerical_qutrit_error}, and gets suppressed when the drive is detuned from these frequencies. 
We can reach lower errors, either by reducing the crosstalk coupling $X$, or by increasing the detuning of the qubit frequency from the drive line frequency.
For instance, achieving a crosstalk-induced error of $10^{-4}$ requires $X<-\SI{42}{dB}$ when the drive and the qubit are on resonance. When the crosstalk is higher, e.g. $X=\SI{-10}{dB}$, one has to increase the detuning $\Delta_\mathrm{dq}/2\pi > \SI{80}{MHz}$ to reach an error below $10^{-4}$.

\begin{figure}[t!]
    \includegraphics{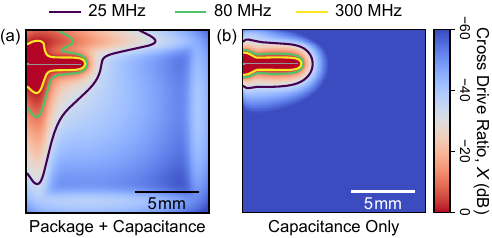}
    \caption{\label{fig:fig15_layout_constraint}
    {\bf Proximity-induced crosstalk over the sample.}
    Simulated cross-drive ratio $X$ at $\SI{5}{GHz}$ across the full sample area, assuming the presence of {\bf (a)} both packaged-mediated and capacitive crosstalk, or {\bf (b)} capacitive crosstalk only. The contours highlights $10^{-4}$ crosstalk-induced single-qubit error with drive-qubit detuning $\Delta_\mathrm{dq} =$ \SI{25}{MHz} (purple), \SI{80}{MHz} (green), and \SI{300}{MHz} (yellow).
    }
\end{figure}

In the presence of proximity-induced crosstalk, the physical separation between qubits and drive lines must be carefully selected to achieve sufficiently low crosstalk-induced errors.
We consider a \SI{5}{GHz} drive signal applied to a drive line of the proximity-induced crosstalk test sample.
According to the simulated charge distribution, we calculate the cross-drive ratio $X$ over the sample, see Fig.~\ref{fig:fig15_layout_constraint}a.
We highlight the contours for $10^{-4}$ crosstalk-induced error corresponding to detuning $\Delta_\mathrm{dq}/2\pi=300$, 80, and \SI{25}{MHz}.
The area within the contour covers $4.2\%$, $9.0\%$, and $24.8\%$ of the full device. The cross-coupled qubit must be positioned outside these areas to achieve an error lower than $10^{-4}$. 

For comparison, we calculate the contours assuming only the direct capacitive crosstalk.
Under this assumption, the area enclosed by the three contours covers $2.4\%$, $3.6\%$, and $8.5\%$ of the full device, see Fig.~\ref{fig:fig15_layout_constraint}b, with approximately a three-fold reduction compared to the case including the package-mediated crosstalk.
With the stray field in the sample package, the microwave crosstalk is decreasing less efficiently when positioning the qubit at a larger distance from the drive line, posing more constraints on the placement of the qubits, especially on large-scale devices where the space is limited and the frequency crowding is more difficult to avoid.
The package-induced crosstalk may be reduced by designing chip-to-PCB transitions that couples less to the package volume.

\section{Case study: characterizing crosstalk on a 17-qubit planar superconducting quantum device} \label{app:crosstalk_on_s17}

We apply our crosstalk models to the leading crosstalk instances on a 17-qubit superconducting quantum processor used for quantum error correction experiments in the surface code~\cite{Besedin2026}. The transmon qubits are positioned on the vertices of a two-dimensional rectangular grid with static nearest-neighboring coupling mediated by capacitively coupled CPWs.
The qubit frequencies are arranged in two frequency bands, which results in approximately \SI{1.5}{GHz} detuning between neighboring qubits.

\begin{figure}[t!]
    \includegraphics{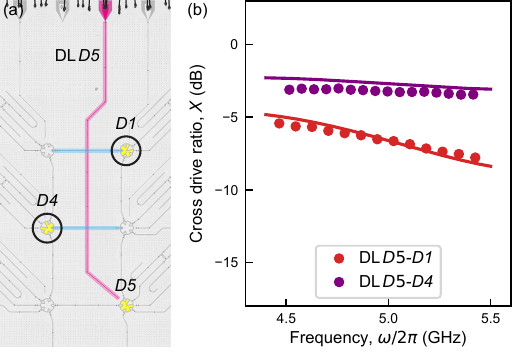}
    \caption{\label{fig:fig16_s17v3p3_crosstalk}
    {\bf Microwave crosstalk characterized on a 17-qubit superconducting quantum processor.} 
    (a) Measured (circles) and simulated (lines) cross-drive ratio $X$ between drive line DL{\it D5} and qubits {\it D1} (red) and  {\it D4} (purple).
    (b) False-colored micrograph of the relevant part of the device, highlighting the microwave drive line DL{\it D5} (pink), the involved transmon qubits {\it D1} and {\it D4} (yellow) and the qubit-qubit couplers (blue).
    } 
\end{figure}

To access qubit $D5$ at the center of the grid, the drive line $\mathrm{DL}\,D5$ crosses two qubit-qubit couplers, see Fig.~\ref{fig:fig16_s17v3p3_crosstalk}a, introducing the strongest crosstalk on the device. 
Qubits $D1$ and $D4$ have their transition frequencies in the same frequency band as qubit $D5$. 
Their operating frequencies therefore need to be chosen sufficiently detuned from that of $D5$.
If this cannot be achieved, coherent cancellation tones may be applied through their respective drive lines to compensate the cross-driving from $\mathrm{DL}\,D5$~\cite{Krinner2022}.

Following the method discussed in Section~\ref{sec:quantifying_microwave_crosstalk} in the main text, we characterize the cross-drive ratio $X$ between $\mathrm{DL}\,D5$ and $D1$, $D4$ as a function of frequency between 4.5 and \SI{5.5}{GHz}.
All three mechanisms described in the main text are present in the studied device.
We consider the combination of all described mechanisms, and find good agreement between the simulations (lines) and the measurements (circles), see Fig.~\ref{fig:fig16_s17v3p3_crosstalk}b.
In this device, the microwave crosstalk is dominated by the crossover.
As the frequency increases, the voltage node move closer to the relevant crossovers, which leads to a decrease in $X$.
The quantitative agreement between the model and the measurement on an intermediate-scale device further supports the improved understanding of microwave crosstalk mechanisms in typical planar superconducting devices.
\bibliographystyle{apsrev4-2-title-etal}

\end{document}